\documentclass[12pt,english,aps,showpacs,superscriptaddress]{revtex4}
\usepackage[T1]{fontenc}
\usepackage{graphicx}

\makeatletter

\usepackage{epsfig}

\usepackage{babel}
\makeatother
\begin{document}
\thispagestyle{empty}

\title{Measurement of the $e^+e^- \to K^+K^-$  cross section
in the energy range $\sqrt{s}$ = 1.04--1.38~GeV  
with the SND detector at VEPP-2M  $e^+e^-$ collider}

\begin{abstract}
The cross section of the process $e^+e^- \to K^+K^-$ was measured in 
the energy range $\sqrt{s}$ = 1.04--1.38~GeV in the SND experiment 
at the VEPP-2M  $e^+e^-$ collider. The measured cross section is 
described by the model of
Vector Meson Dominance with contributions from the light vector mesons 
$\rho$, $\omega$, $\phi$ and their lowest excitations. The mean statistical 
accuracy of the measurement is 4.4~$\%$, and the systematic uncertainty is 
5.2~$\%$.
\end{abstract}


\author{M.~N.~Achasov}

\affiliation{Budker Institute of Nuclear Physics, Novosibirsk, 630090, Russia}

\affiliation{Novosibirsk State University, Novosibirsk, 630090, Russia}

\author{K.~I.~Beloborodov}

\affiliation{Budker Institute of Nuclear Physics, Novosibirsk, 630090, Russia}

\affiliation{Novosibirsk State University, Novosibirsk, 630090, Russia}

\author{A.~V.~Berdyugin}

\affiliation{Budker Institute of Nuclear Physics, Novosibirsk, 630090, Russia}

\affiliation{Novosibirsk State University, Novosibirsk, 630090, Russia}

\author{A.~G.~Bogdanchikov}

\affiliation{Budker Institute of Nuclear Physics, Novosibirsk, 630090, Russia}

\author{A.~D.~Bukin}

\affiliation{Budker Institute of Nuclear Physics, Novosibirsk, 630090, Russia}

\affiliation{Novosibirsk State University, Novosibirsk, 630090, Russia}

\author{D.~A.~Bukin}

\affiliation{Budker Institute of Nuclear Physics, Novosibirsk, 630090, Russia}

\author{T.~V.~Dimova}

\affiliation{Budker Institute of Nuclear Physics, Novosibirsk, 630090, Russia}

\affiliation{Novosibirsk State University, Novosibirsk, 630090, Russia}

\author{V.~P.~Druzhinin}

\affiliation{Budker Institute of Nuclear Physics, Novosibirsk, 630090, Russia}

\affiliation{Novosibirsk State University, Novosibirsk, 630090, Russia}

\author{V.~B.~Golubev}

\affiliation{Budker Institute of Nuclear Physics, Novosibirsk, 630090, Russia}

\affiliation{Novosibirsk State University, Novosibirsk, 630090, Russia}

\author{I.~A.~Koop}

\affiliation{Budker Institute of Nuclear Physics, Novosibirsk, 630090, Russia}

\affiliation{Novosibirsk State University, Novosibirsk, 630090, Russia}

\author{A.~A.~Korol}

\affiliation{Budker Institute of Nuclear Physics, Novosibirsk, 630090, Russia}

\author{S.~V.~Koshuba}

\affiliation{Budker Institute of Nuclear Physics, Novosibirsk, 630090, Russia}

\author{E.~V.~Pakhtusova}

\affiliation{Budker Institute of Nuclear Physics, Novosibirsk, 630090, Russia}

\author{E.~A.~Perevedentsev}

\affiliation{Budker Institute of Nuclear Physics, Novosibirsk, 630090, Russia}

\author{S.~I.~Serednyakov}

\affiliation{Budker Institute of Nuclear Physics, Novosibirsk, 630090, Russia}

\affiliation{Novosibirsk State University, Novosibirsk, 630090, Russia}

\author{Yu.~M.~Shatunov}

\affiliation{Budker Institute of Nuclear Physics, Novosibirsk, 630090, Russia}

\affiliation{Novosibirsk State University, Novosibirsk, 630090, Russia}

\author{Z.~K.~Silagadze}

\affiliation{Budker Institute of Nuclear Physics, Novosibirsk, 630090, Russia}

\affiliation{Novosibirsk State University, Novosibirsk, 630090, Russia}

\author{A.~N.~Skrinsky}

\affiliation{Budker Institute of Nuclear Physics, Novosibirsk, 630090, Russia}

\author{Yu.~V.~Usov}

\affiliation{Budker Institute of Nuclear Physics, Novosibirsk, 630090, Russia}

\author{A.~V.~Vasiljev}

\affiliation{Budker Institute of Nuclear Physics, Novosibirsk, 630090, Russia}

\affiliation{Novosibirsk State University, Novosibirsk, 630090, Russia}


\maketitle

\section{Introduction}

The experimental study of the process $e^+e^- \to K^+K^-$ at energies 
$\geq 1$~GeV is of interest for a number of reasons. First, this is one 
of the processes in which the excited states of the light vector
mesons  whose characteristics until now have not been reliably  
established, should manifest themselves. Second, the cross section of 
this process is determined by the form factor of charged kaon, 
whose knowledge together with the data on the neutral kaon form
factor, measured  in the reaction $e^+e^- \to K_SK_L$~\cite{sndkskl} 
at the same energies, makes possible a calculation of the isovector 
and isoscalar form factors of kaons and determination of  
the parameters of the excited states of light vector mesons. 
Third, the isovector form factor of a kaon obtained in $e^+e^-$ annihilation 
can be used to test conservation of vector current by using 
experimental data on the $\tau$ lepton decay modes with kaons~\cite{bruch}. 
And finally, the process production of charged kaon pairs contributes 
to the total hadron production cross section and its accurate knowledge 
is interesting for the Standard Model tests, for example, for the 
measurement of the muon anomalous magnetic moment.

The first observation of pair production of charged kaons in $e^+e^-$ 
interactions in the energy region above the $\phi$ meson resonance 
was performed at Novosibirsk collider VEPP-2 in 1970~\cite{onuchin}. 
Subsequently the study of $e^+e^- \to K^+K^-$ was conducted in 
experiments~\cite{grosdidier,olya,cmd1}. The most precise 
measurement of the cross section (10\% statistical and 10$\%$ systematic 
uncertainties) was performed with the OLYA detector
at the VEPP-2M collider~\cite{olya,lelchuk}. 
The urgency of the charged kaon form factor 
measurement is also motivated by a recent 
announcement~\cite{kkresonance} about the observation 
of a wide resonance structure in the $K^+K^-$ invariant mass spectrum near 
1.5~GeV in the decay $J/\psi \to K^+K^-\pi^0$.

In the present work we report the results of the measurement
of the $e^+e^- \to K^+K^-$ cross section in experiments with the 
SND detector at the VEPP-2M collider in the center-of-mass energy range
$\sqrt{s}$ from $1.04$ to $1.38$~GeV.

\section{ The SND detector and experiment}

In 1995--2000 experiments with the SND detector~\cite{SND} were 
carried out at the VEPP-2M  $e^+e^-$
collider ~\cite{VEPP2M} in the energy range from 
0.36 to 1.38~GeV.  The basic part of the SND detector is the three-layer 
electromagnetic calorimeter consisting of 1632 NaI(Tl) crystals. 
The energy and angular resolutions of the calorimeter depend on 
photon energy: $\sigma_{E}/E(\%) = 4.2\%/\sqrt[4]{E(GeV)}$ and
 $\sigma_{\varphi,\theta}=0.82^{\circ}/\sqrt{E(GeV)}\oplus0.63^{\circ}$.
Charged-particle tracking is provided by two coaxial cylindrical drift 
chambers. The angular resolution is 0.5$^{\circ}$ and 2$^{\circ}$ 
for the azimuthal and polar angles, respectively. Muon identification 
is provided by the muon system consisting of two layers of streamer tubes
and a layer of plastic scintillation counters. The solid angle coverage 
of the SND detector is 90\% of 4$\pi$.

In the present work the data of two scans taken in 1997    
(from 0.98 to 1.38~GeV) and one scan taken in 1999 
(from 1.04 to 1.34~GeV) are used. 
The integrated luminosity of these experiments measured using $e^+e^-$
elastic scattering is 6.7 pb$^{-1}$.
The energy spread at each energy 
point does not exceed 1~MeV and is taken into account in the data analysis.

\section{ Event selection }
To select events of the process 
\begin{equation}
\label{k1}
e^+e^- \to K^+K^-,
\end{equation}
we require the presence of two collinear charged particles originating
from the $e^+e^-$ interaction point.
A deviation from collinearity ($\Delta\varphi$) in the plane 
perpendicular to the beam direction is determined  
by multiple scattering in the detector material in front of  
the tracking system and depends on the kaon energy. The 
following conditions on this parameter were required: \\
 $|\Delta\varphi| \leq 15^\circ$ for   $\sqrt{s}~<~1.08$~GeV, \\
 $|\Delta\varphi| \leq 8^\circ$ for $1.08$~GeV~$\leq~\sqrt{s}~<~1.1$~GeV,\\
 $|\Delta\varphi| \leq 5^\circ$ for $\sqrt{s}~\geq~1.1$~GeV.\\
A deviation from collinearity ($\Delta\theta$) in the plane passing 
through the beam is additionally affected by  
radiation from the initial  $e^+e^-$ state.
To limit  the energy of radiated photons,
the condition $|\Delta\theta| \leq 10^\circ$ was used. It was 
required for both tracks that $r_i \leq 0.3$~cm and $|z_i|~\leq~10$~cm,
where $r_i$ is the minimum distance from the track to the beam axis 
and $z_i$ is the coordinate of the particle production point along 
the beam axis.
The number of neutral particles detected in the event was not limited.

Basic background processes satisfying the  
selection criteria are processes with two collinear particles in the final 
state:
\begin{equation}
\label{pi1}
e^+e^- \to \pi^+\pi^-
\end{equation}
\begin{equation}
\label{e1}
  e^+e^- \to e^+e^-
\end{equation}
\begin{equation}
\label{mu1}
  e^+e^- \to \mu^+\mu^-
\end{equation}  

To suppress the background from the process (\ref{e1}), 
we required that the energy deposition of the charged particles in
the calorimeter be less than 0.7~$\sqrt{s}$ and their polar
angles $\theta_i$ be in the range $ 50^\circ < \theta_i < 130^\circ$.
The substantial part of the background from the process (\ref{e1}) is caused 
by particle hittings into 
the dead calorimeter counters.
Such events, whose fraction is about 2\%, were excluded from the analysis.

To suppress background from the process  (\ref{mu1}) and  reject 
the cosmic-ray background, a signal from the muon  
system was used. The requirement of the absence of this signal decreased the 
background from the process (\ref{mu1}) by approximately 2 orders of 
magnitude.

With these criteria 136532 events were selected
in the energy interval from $1.04$ to $1.38$~GeV.
 
\section{ Background processes} 
 For the additional suppression of the background from the processes 
 (\ref{pi1})--(\ref{mu1}), we used a difference in the energy deposition
profile in the calorimeter layers for particles of the signal 
and background processes. Efficient   
$K-\pi$ separation provided by the calorimeter is possible in the 
energy range under study because of the substantial difference between 
kaons and pions  in the ionization losses and in the penetration depths. 
The special separation parameters were created
with the aid  of the neural network approach~\cite{NEURON}. 
Monte-Carlo simulated events were used for the network training.

\begin{figure}[htbp]
\includegraphics[width=0.75\textwidth]{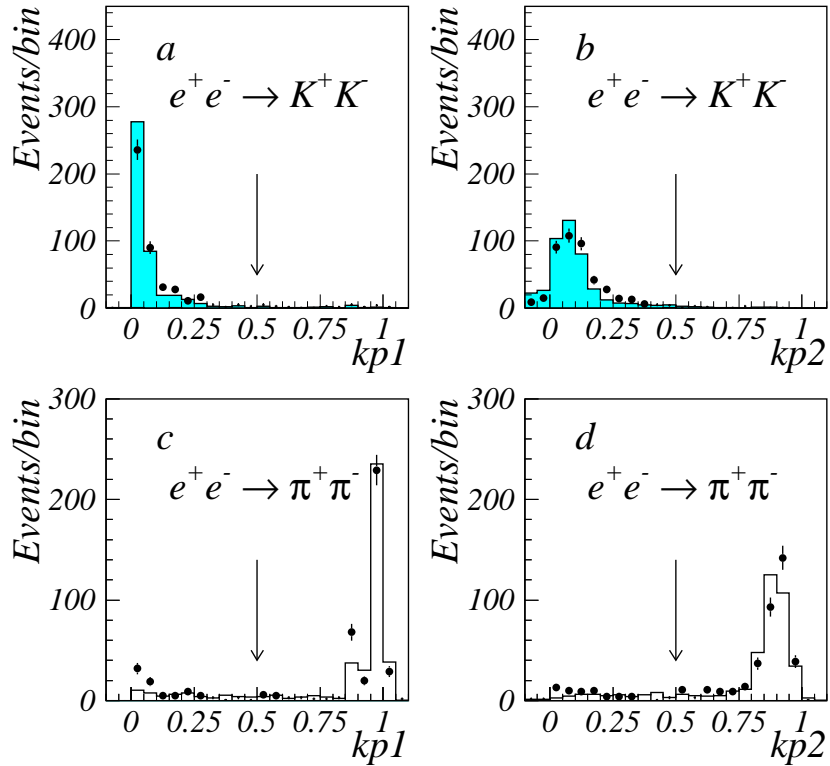}
\caption{ \label{kp1}
Distributions of the $K-\pi$ separation parameters for 
data (points with error bars) and simulated (histogram) events of 
the processes (\ref{k1}) and  (\ref{pi1})  
for the energy point $\sqrt{s}$ = 1.2~GeV.
(a),(c) -- the first particle, (b),(d) -- the second particle.  
The cuts to select $e^+e^- \to K^+K^-$ events  are indicated by arrows.
}
\end{figure}

\begin{figure}[htbp]
\includegraphics[width=0.75\textwidth]{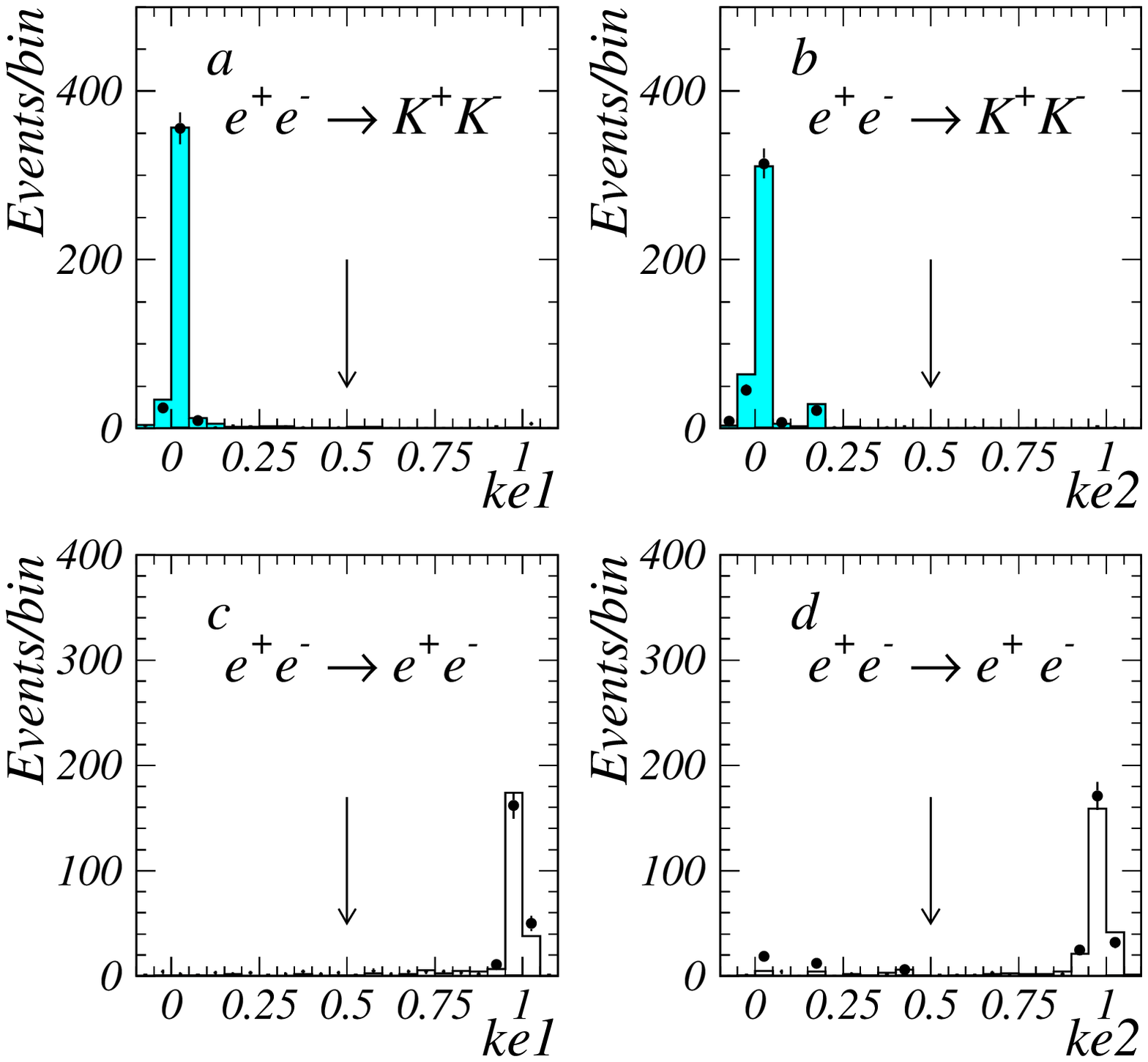}
\caption{ \label{ke1}
Distributions of the $K-e$ separation parameters for 
data (points with error bars) and simulated (histogram) events of 
the processes (\ref{k1}) and  (\ref{e1})  
for the energy point $\sqrt{s}$ = 1.2~GeV.
(a),(c) -- the first particle, (b),(d) -- the second particle.  
The cuts to select $e^+e^- \to K^+K^-$ events  are indicated by arrows.
}
\end{figure}

For each detected particle in the event 
parameters were created to separate kaons and 
pions $(kp1, kp2)$, kaons and electrons $(ke1, ke2)$, kaons and  muons  
$(km1, km2)$, pions and muons of $(pm1, pm2)$, pions and electrons
$(pe1, pe2)$. The separation parameters were obtained  for each 
energy point in the experiment.      
Distributions of the separation parameters for the data and simulated
events of the signal and background processes are shown in Fig.\ref{kp1} and 
\ref{ke1} for the energy point $\sqrt{s}$=1.2~GeV. Particles are numbered 
according to the value of the energy deposition in the calorimeter, the first 
particle being the one with the greatest energy deposition.
To obtain the data distribution for the first (second) particle of 
the particular process
the tight cuts on the separation parameters of the second (first)
particle were used. 

For the selection of $e^+e^- \to K^+K^-$ events, the following 
restrictions on the separation parameters of the particles were used:
\begin{equation}
\label{kk}
\begin{array}{l}
kp1 < 0.5,\;kp2<0.5,  \\
ke1 < 0.5,\;ke2<0.5.
\end{array}
\end{equation}

\begin{figure}[htbp]
\includegraphics[width=0.75\textwidth]{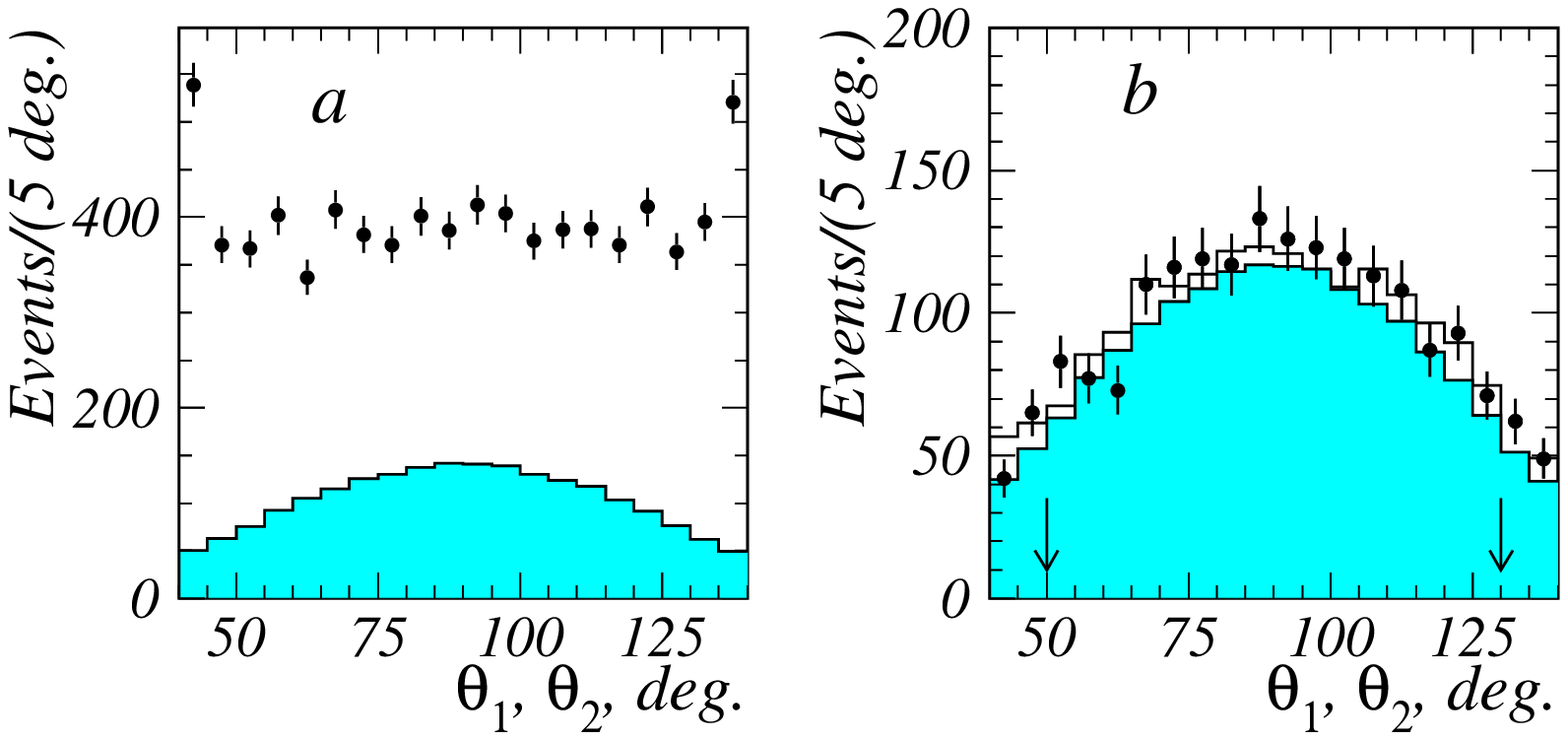}
\caption{ \label{tet1299600kk_1}
The polar angle distribution for data events (points with error bars)
before (a) and after (b) applying the selection criteria (\ref{kk}).
The shaded histogram shows the distribution for simulated $e^+e^- \to K^+K^-$
events. The hollow histogram is the sum of simulated distributions
for the signal and background processes.   
Arrows indicate the selection boundaries 
on the angles $\theta_1$, $\theta_2$. Data are for the energy $\sqrt{s}$=
1.24~GeV.
}
\end{figure}

The role of conditions (\ref{kk}) in the suppression of the background is  
demonstrated in Fig.\ref{tet1299600kk_1}, where the polar angle 
distribution for data events is shown  
before and after applying the selection criteria (\ref{kk}).
While without the selection cuts (\ref{kk}) the        
background  exceeded the signal by a factor of more than three, with the 
selection cuts it was reduced to a few percent only.

After applying the conditions (\ref{kk}), the contribution from the 
background process (\ref{mu1}) was found to be negligible 
(less than 0.1\% of the number of selected $K^+K^-$ events ). 
The contributions of background 
processes (\ref{pi1}) and (\ref{e1}) were determined from the simulation.  
Normalizing coefficients for the conversion of the number of events in 
simulation to the expected number of events in the data were 
determined in the following way. Using the tight conditions on the 
separation parameters, three classes of events were selected
in which one of the processes (\ref{k1})--(\ref{e1}) 
dominates. 
The numbers of data events in these classes ($N^{KK}_{exp}$, 
$N^{\pi\pi}_{exp}$ and $N^{ee}_{exp}$) are related to the numbers of
simulated events of the processes $e^+e^- \to f, \; f = K^+K^-, \; \pi^+\pi^-,
\; e^+e^-$ assigned to the definite class ($N^{KK}_{mc,f}$, $N^{\pi\pi}_
{mc,f}$ and $N^{ee}_{mc,f}$) in the following way 
\begin{equation}
\label{eq1}
\begin{array}{l}
N_{exp}^{KK} = \sum\limits_{f} N_{mc,f}^{KK}\cdot k_f, \\
N_{exp}^{\pi\pi} = \sum\limits_{f}N_{mc,f}^{\pi\pi}\cdot k_f, \\
N_{exp}^{ee} = \sum\limits_{f}N_{mc,f}^{ee}\cdot k_f, \\
\end{array}
\end{equation}
where $k_f$ are 
normalizing coefficients to be found. A linear 
system of equations (\ref{eq1}) allows to find coefficients  $k_{K^+K^-}$, 
$k_{\pi^+\pi^-}$, $k_{e^+e^-}$  for each energy point. While determining
$N^{KK}_{exp}$, $N^{\pi\pi}_{exp}$ and $N^{ee}_{exp}$, contributions from
the beam and non-collinear backgrounds were taken into account and subtracted
by the method described below. 
The origin of beam background is electron or positron collisions 
with residual gas in the accelerator beam pipe  near the interaction point.
Background from the process $e^+e^- \to 
\mu^+\mu^-$ in the pion class was suppressed using the $\pi-\mu$ 
separation parameters. The obtained coefficients $k_{\pi^+\pi^-}$ and 
$k_{e^+e^-}$ were used for estimation of the pion and electron
background in the event sample selected with standard
criteria (\ref{kk}).

Besides processes with the collinear charged particles, other potential
background sources are $e^+e^-$ annihilation to three or more particles in the
final state
\begin{equation}
\begin{array}{l}
\label{omegapi}
  e^+e^- \to \pi^+\pi^-\pi^0\pi^0, \\
  e^+e^- \to \pi^+\pi^-\pi^+\pi^-, \\
  e^+e^- \to \pi^+\pi^-\pi^0,
\end{array}
\end{equation}
and the beam related background.

\begin{figure}[htbp]
\includegraphics[width=0.96\textwidth]{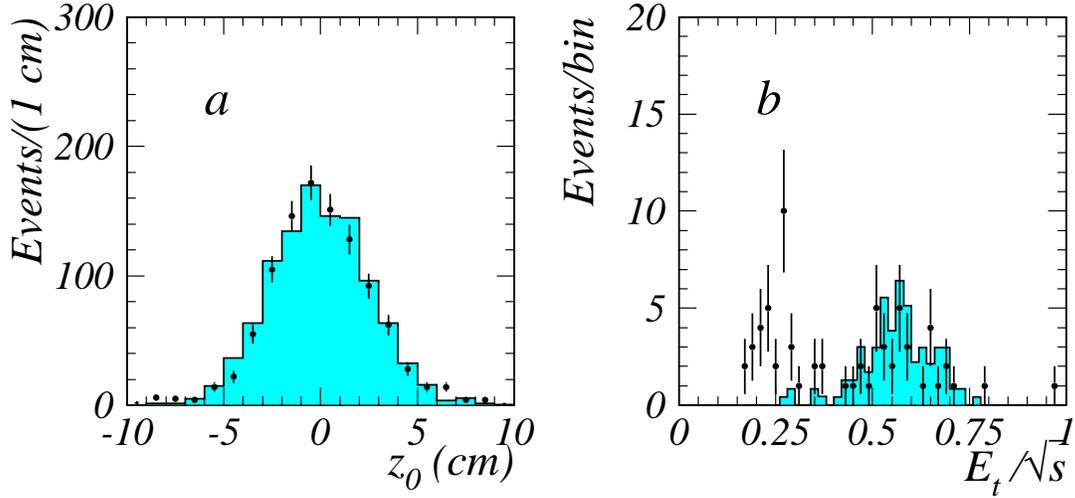}
\caption{ \label{kaon680z12eton}
(a) Distribution of the event-vertex coordinate $z_0= (z_1 + z_2)/2$ 
for selected data events (points with error bars) and simulated
signal and background events (shaded histogram), 
$\sqrt{s}$=1.36~GeV. (b) Distribution of the normalized 
energy deposition in the calorimeter $E_{t}$/$\sqrt{s}$ for events with 
5~cm $ \leq|z_0|\leq $ 10~cm. Points with error bars  represent the data. 
The shaded histogram is the simulation of the 
$e^+e^- \to K^+K^-$, $e^+e^- \to \pi^+\pi^-$, and $e^+e^- \to e^+e^-$.
}
\end{figure}

The distributions of the event-vertex coordinate $z_0= (z_1 + z_2)/2$
for the selected data events and simulated events of the processes 
(\ref{k1})--(\ref{e1}) is shown 
in Fig.\ref{kaon680z12eton}(a) for the energy point $\sqrt{s}$=1.36~GeV.
\noindent
Most of the $e^+e^-$ annihilation events are located in the range 
$|z_0|\leq $5~cm. The beam background becomes  
significant in the region $|z_0| > $5~cm.  
The  distribution of the normalized energy deposition in the calorimeter 
$E_{t}$/$\sqrt{s}$ for the events  with 5~cm $ \leq|z_0|\leq $ 10~cm is shown
in Fig.\ref{kaon680z12eton}(b). Events  with     
$E_{t}$/$\sqrt{s}~\leq $~0.35 are mainly from the beam background.
They are uniformly distributed over $ |z_0| $ 
while their 
distances  from the track to the beam axis
are concentrated near zero.
The contribution of the beam background to the selected data events was 
estimated as twice the number of events with 5~cm $ \leq|z_0|\leq $ 10~cm and 
$E_{t}$/$\sqrt{s} \leq $ 0.35 minus the calculated number of events of the
processes (\ref{k1})--(\ref{e1}). 

\begin{figure}[htbp]
\includegraphics[width=0.96\textwidth]{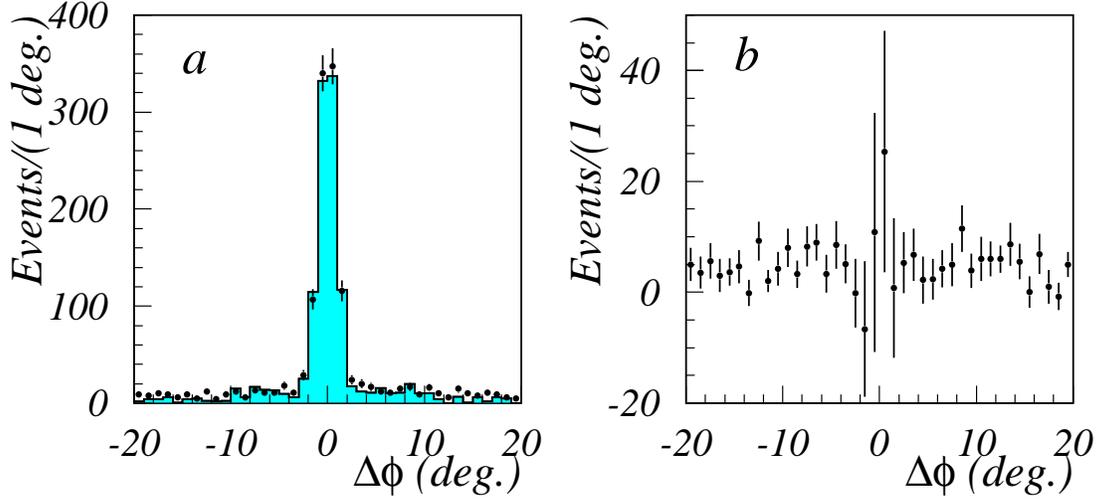}
\caption{ \label{kaon680035g}
(a) The $\Delta\varphi$ distribution for selected data events 
(points with error bars). The shaded histogram shows the combined 
distribution for simulated events
of the processes (\ref{k1})--(\ref{e1})   
and events of the beam background; $\sqrt{s}$=1.36~GeV. 
(b) The difference between the distributions shown in (a).
}
\end{figure}

The $\Delta\varphi$ distributions for selected data events 
and simulated events of the processes (\ref{k1})--(\ref{e1}) 
and of the beam background 
are shown in Fig.\ref{kaon680035g}(a). The difference between 
the data and simulation distributions is shown in  Fig.~\ref{kaon680035g}(b).
It is seen that the remained background is uniformly distributed over the 
$\Delta\varphi$, as expected for the processes (\ref{omegapi}).
The number of background events of the processes  (\ref{omegapi}) 
was estimated as the number of events with
$\Delta\varphi_0 < |\Delta\varphi| < 2\Delta\varphi_0$
minus expected number of events of the processes
(\ref{k1})--(\ref{e1}) and beam background. 
Here $\Delta\varphi_0$ equals  $5^\circ$, $8^\circ$ or $15^\circ$ 
depending on the energy point.

\begin{figure}[htbp]
\begin{minipage}[t]{0.44\textwidth}
\includegraphics[width=\textwidth]{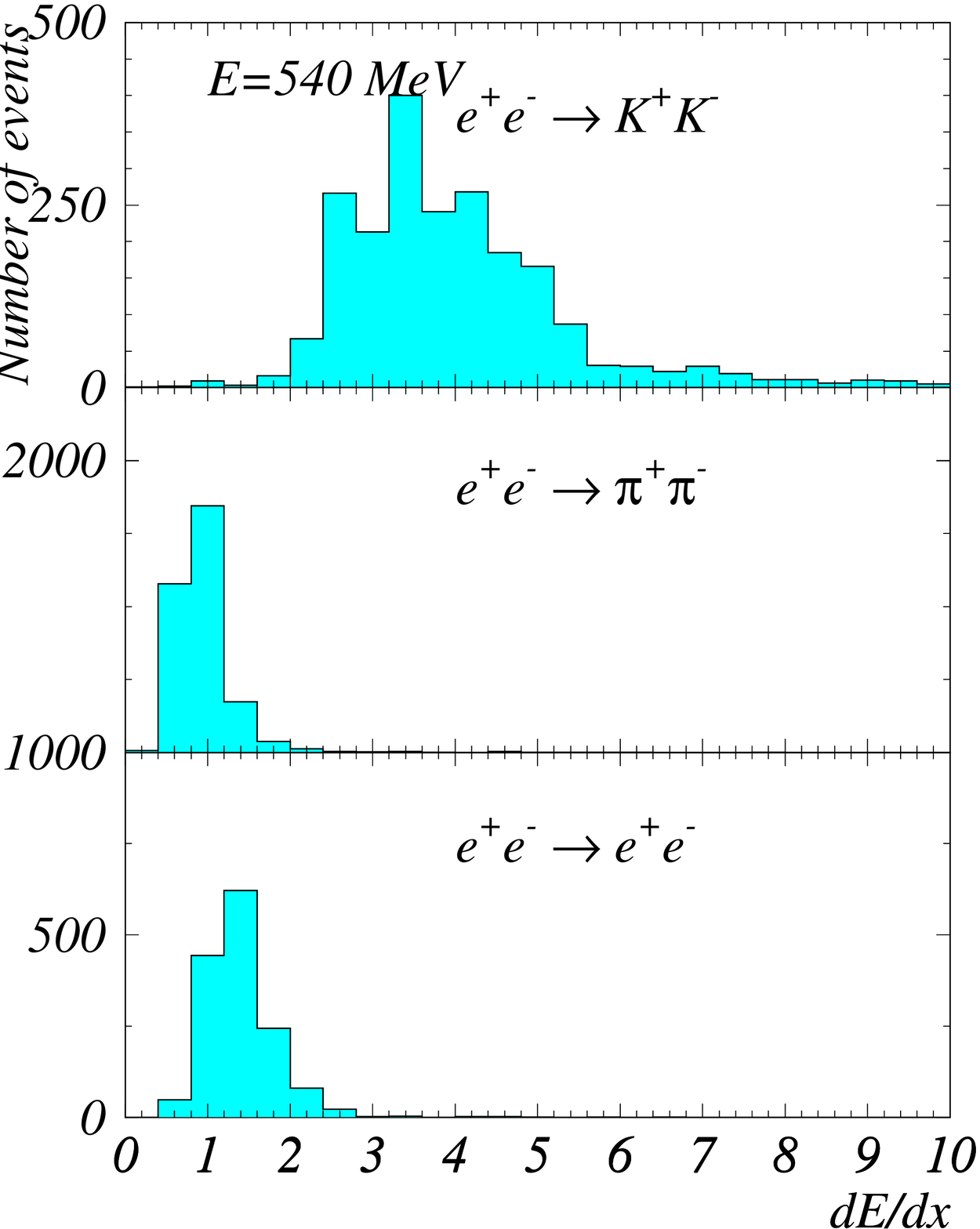}
\end{minipage}
\hspace{-2mm}
\begin{minipage}[t]{0.44\textwidth}
\includegraphics[width=\textwidth]{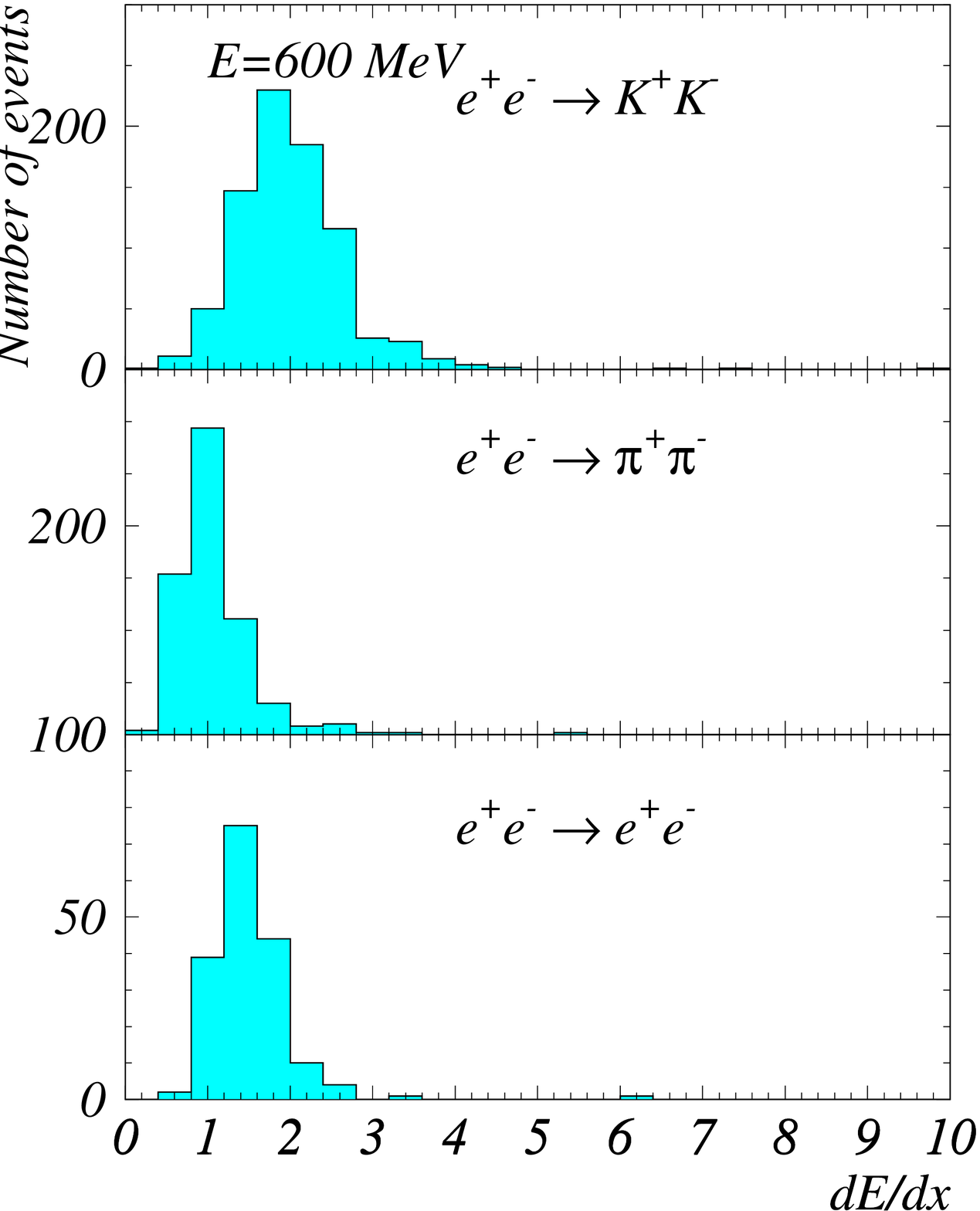}
\end{minipage}\\[-2mm]
\caption{ \label{dedx}
The distributions of the ionization losses $(dE/dx)_1$ of the first 
particle for data events of the 
processes  $e^+e^- \to K^+K^-$, $e^+e^- \to \pi^+\pi^-$ and 
$e^+e^- \to e^+e^-$. (a) $\sqrt{s}$=1.08~GeV,
(b) $\sqrt{s}$=1.20~GeV.
}
\end{figure}

Additional suppression of the background,  both collinear and non-collinear, 
is provided by cuts on the ionization losses  ( normalized to those of the
minimum  ionizing particle ) $(dE/dx)$ in the drift chambers. 
The $(dE/dx)$ distributions of the first particle for the data events of the
processes $e^+e^- \to K^+K^-$, $e^+e^- \to \pi^+\pi^-$ and $e^+e^- \to e^+e^-$ 
are shown in Fig.~\ref{dedx} for the energy points  $\sqrt{s}$=1.08~GeV
and 1.20~GeV.
For events with  $\sqrt{s} <1.20$~GeV, we have used the following cuts on this 
parameter:\\
$\sqrt{s} \leq 1.11$~GeV:~~~~~~~~~~~~~$(dE/dx)_1~>~2$; \\
$\sqrt{s}  =1.12\div1.18$~GeV: ~~$(dE/dx)_1~>~1.5$;\\
$\sqrt{s} =1.19\div1.20$~GeV:   ~~$(dE/dx)_1~>~1$.\\

\begin{figure}[htbp]
\includegraphics[width=0.68\textwidth]{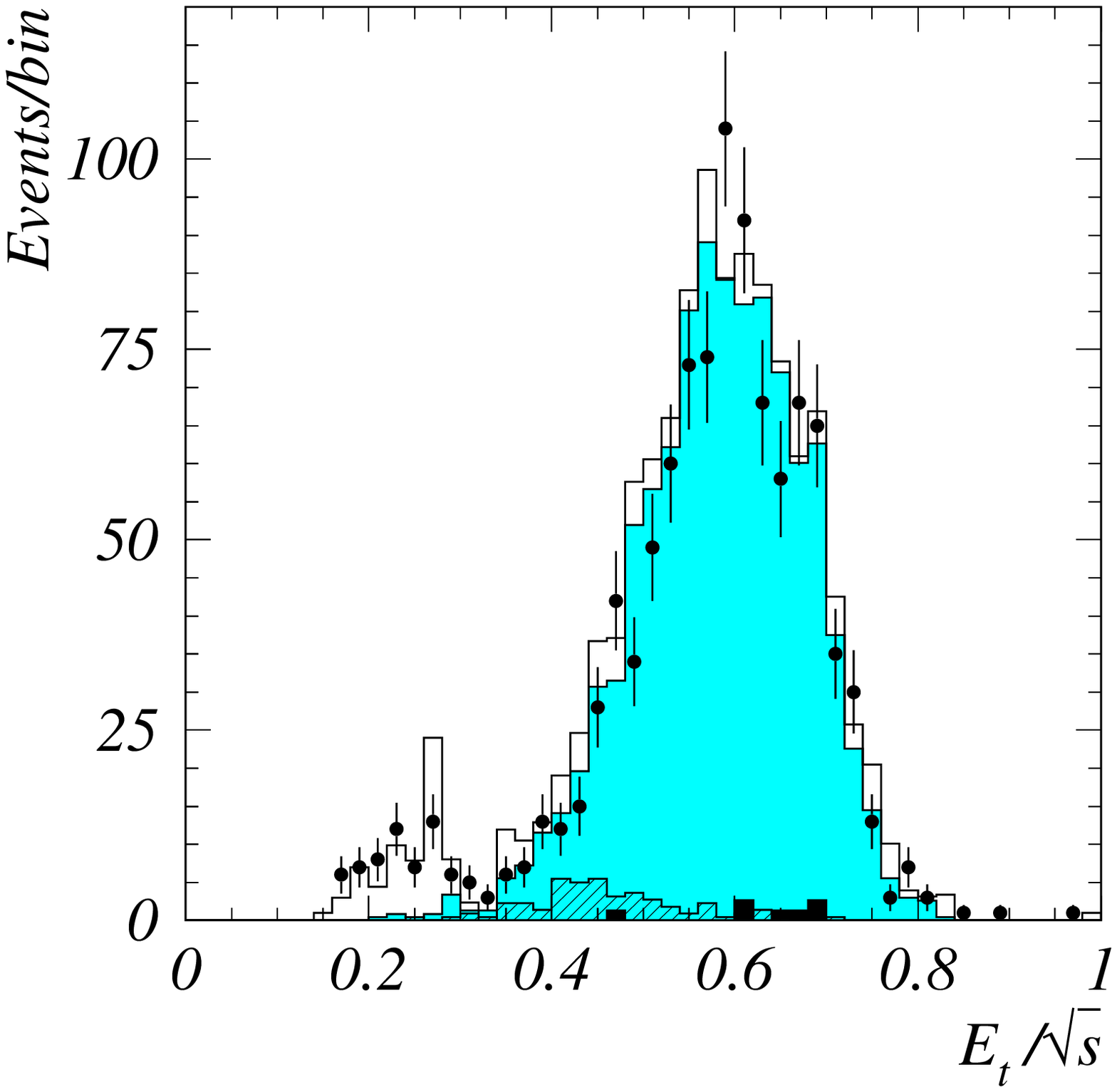}
\caption{ \label{kaon680035_1}
Distributions of the normalized energy deposition in the calorimeter
$E_{t}$/$\sqrt{s}$ at $\sqrt{s}$=1.36~GeV. The shaded, hashed, and solid
histograms are simulations of the processes $e^+e^- \to K^+K^-$,
$e^+e^- \to \pi^+\pi^-$, and $e^+e^- \to e^+e^-$, respectively. 
The hollow histogram shows
the total contribution from the process $e^+e^- \to K^+K^-$ and from all
background processes including the beam background. 
}
\end{figure}

The distributions of the normalized energy deposition in the calorimeter 
for data and simulated events at $\sqrt{s}$=1.36~GeV are shown 
in Fig.\ref{kaon680035_1}. It is evident that at this energy the 
beam background can be efficiently eliminated by the condition  
$E_{t}$/$\sqrt{s} \geq $ 0.3 without essential loss of efficiency to 
the process under study. This additional condition was used for the energy 
points with  $\sqrt{s} \geq 1.22$~GeV.

After applying the above mentioned selection criteria and background 
subtraction, 54402 events of the process  $e^+e^- \to K^+K^-$ were found.
The numbers of events found for the signal (\ref{k1}) and background processes 
are presented in Table~\ref{tabnev} for some energies.
Errors in the number of kaons include uncertainties due to the 
statistical errors
of the $k_{\pi^+\pi^-}$ and $k_{e^+e^-}$ coefficients and  
the subtraction of non-collinear and beam backgrounds.
\begin{table}[htb]
\begin{center}
\caption{ \label{tabnev}
$e^+e^-$ center-of-mass energy ($\sqrt{s}$),
the number of selected $e^+e^- \to K^+K^-$ events ($N(K^+K^-)$) with
subtracted background, 
calculated numbers of background events from $e^+e^- \to \pi^+\pi^-$
($\bar{N}_{\pi^+\pi^-}$) and $e^+e^- \to e^+e^-$ ($\bar{N}_{e^+e^-}$),
numbers of beam background events ($N_1$) and events of
the $e^+e^-$ annihilation into the final states
with two non-collinear charged particles 
($N_2$).\\
} 
\begin{tabular}{|c|c|c|c|c|c|}
\hline
$\sqrt{s}$, GeV & $N(K^+K^-)$ & $\bar{N}_{\pi^+\pi^-}$ &
$\bar{N}_{e^+e^-}$
& $N_1$ & $N_2$   \\
 \hline
1.04  & $1347.4\pm38.5$ & $0.1 \pm 0.1 $ & $ \leq 0.07 $ & $10.3 \pm 6.8$ 
& $10.2 \pm 4.5$  \\
1.10  & $2042.0 \pm 49.3$ & $3.0 \pm 1.3$ & $ \leq 0.9$ & $31.5 \pm 9.9$ 
& $ 30.5 \pm  8.1 $\\
1.15  & $281.7\pm 17.4$ & $3.1\pm2.4$ & $ 0.3\pm0.3$ & $3.3\pm 0.5$
& $4.0 \pm 3.5$ \\ 
1.20 & $1600.2\pm 42.5$ & $38.7 \pm 2.3$  & $\leq 1.1$   & $ 28.3 \pm8.0$ &
$25.7 \pm  7.4$ \\ 
1.25 & $697.7\pm 28.0$ & $13.1 \pm 2.5$ & $5.3 \pm 2.2$ & - & $14.0\pm4.9$\\ 
1.30 & $792.9 \pm 30.7$ & $32.0\pm 2.3$& $7.6 \pm 2.8$ & - &$41.2 \pm7.3$\\
1.35 &$718.7 \pm  32.9$ & $38.4\pm 14.3$ & $9.0\pm 4.1$ &  - & $ 29.0 \pm 7.9
$\\ 
1.38 &$1127.3 \pm 36.0$ & $41.3\pm3.2$  & $19.4\pm4.7$  & - & $ 28.1 \pm  6.7
$ \\ 
\hline
\end{tabular}
\end{center}
\end{table}
\noindent
The contribution of background processes depends on energy and 
does not exceed 5\% and 2\% of the number of $K^+K^-$ events for the
processes (\ref{pi1}) and (\ref{e1}), respectively. 
For beam background and background from the processes (\ref{omegapi}) 
it is less than 4\% and 5\%, respectively.
 
To estimate a possible systematic uncertainty due to subtraction of
the $e^+e^- \to \pi^+\pi^-$ background we performed its independent
estimation using the difference in the ionization 
losses of kaons and pions in the drift chambers (Fig.~\ref{dedx}).
For the 1999 energy scan, from the ratio of the number of 
events  in the regions with 
$(dE/dx)_1\leq1$ and  $(dE/dx)_1>1$ (or $(dE/dx)_2\leq1$ and $(dE/dx)_2>1$
for $\sqrt{s} <1.20$~GeV) the number of background events $N(\pi^+\pi^-)$
was found to be $ 47\pm 30$ for $\sqrt{s} \leq 1.20$~GeV, and $157 \pm 29$ for 
$\sqrt{s} > 1.20$~GeV. The corresponding numbers of pionic events, 
calculated according to the simulation using the procedure described 
earlier, are $ 58 \pm 2 $ and $117 \pm  6$. Two methods of $e^+e^- \to 
\pi^+\pi^-$ background evaluation give the results which agree with each 
other. Nevertheless, the difference between the two calculations was used as  
an estimate of the accuracy of pionic background determination. 
The corresponding systematic error in the number of events of the process 
$e^+e^-\to K^+K^-$  is 0.1\% for $\sqrt{s} \leq 1.20$~GeV 
($N(K^+K^-) = 11230$), and 1.4\% for $\sqrt{s}>1.20$~GeV ($N(K^+K^-) = 2900$).

The independent estimation of the background from electrons was performed by 
using the difference in the angular distributions for events of the 
$e^+e^- \to e^+e^-$ and the signal process. The ratio of the number of 
events in the regions with $(60^{\circ} < \theta_i < 120^{\circ})$ and
$(50^{\circ}\leq \theta_i < 60^{\circ},\,120^{\circ} < \theta_i \leq 
130^{\circ})$ was used. Events with ($ke1<0.1$, $ke2<0.1$) were rejected
to enhance sensitivity to electrons. 
This condition significantly reduces the
number of $K^+K^-$ events while the efficiency to the background process
remains almost intact. 
The number of background events $N(e^+e^-)$ was found to be
$ 35 \pm 28$ for $\sqrt{s} \leq  1.20$~GeV and $7 \pm  24$
for $\sqrt{s} >  1.20$~GeV. A calculation according to the simulation 
gives 3 and 17 events, respectively. The difference between two estimations,
0.3\% of the number of $K^+K^-$ events for the entire energy range, 
was used as a measure of the systematic uncertainty due to subtraction
of the $e^+e^- \to e^+e^-$ background.  
 
\section{ Detection efficiency }
\begin{figure}[htbp]    
\includegraphics[width=0.68\textwidth]{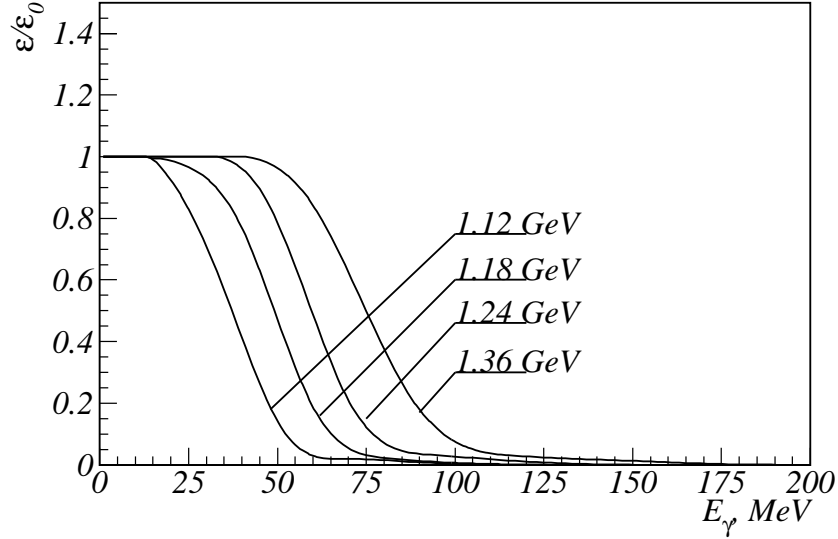}
\caption{ \label{efficiency}
Detection efficiency of the process $e^+e^- \to K^+K^-$ versus the energy of 
the photon emitted by initial particles for  $\sqrt{s}$=1.12, 1.18, 1.24 and 
1.36~GeV.
 }
\end{figure}

The detection efficiency for the process $e^+e^- \to K^+K^-$ was determined 
from Monte-Carlo simulation~\cite{unimod}. In the simulation
the emission of photons by initial particles was taken into 
account~\cite{fadin,bm}, and the detection efficiency was evaluated
as a function of the center-of-mass energy $\sqrt{s}$ and 
the energy $E_{\gamma}$ of the photon radiated by initial particles:
\begin{equation}
\varepsilon(\sqrt{s},E_{\gamma})= \varepsilon_0(\sqrt{s}) 
f(\sqrt{s},E_{\gamma}),
\end{equation}
where $\varepsilon_0(\sqrt{s})$ is the efficiency with no photon emission.
The dependence of $f(\sqrt{s},E_{\gamma})$ on the photon
energy at
$\sqrt{s}$=1.12, 1.18, 1.24, and 1.36~GeV is shown in Fig.\ref{efficiency}. 

The detection efficiency determined from simulation was 
multiplied by the correction factor, which takes into account the 
data-simulation difference in the detector response. 
The total correction factor is the product of correction factors 
related to specific selection condition. 
The correction factor for a given condition was 
calculated as
 \begin{equation}
 \delta_{\varepsilon} = \frac{N_{K^+K^-}(exp)\cdot N^{(1)}_{K^+K^-}(mc)}
          {N^{(1)}_{K^+K^-}(exp)\cdot N_{K^+K^-}(mc)},
 \end{equation}
where $N_{K^+K^-}(exp)$,  $N_{K^+K^-}(mc)$ are the numbers of $K^+K^-$ events
in the data and simulation, respectively, with the condition applied, 
$N^{(1)}_{K^+K^-}(exp)$, $N^{(1)}_{K^+K^-}(mc)$ are 
the numbers of events with the condition removed.
To suppress the background, the conditions on other parameters were 
tightened and 
additional conditions on the ionization losses in the drift chambers were
applied.

To obtain the correction related to the $ \Delta\theta $ cut 
we used events with $|\Delta\theta|\leq 50^{\circ}$. The distribution of 
$ \Delta\theta $ is strongly influenced by the photon emission from the
initial state. Therefore, the experimental dependence of the cross section on 
energy was implemented in the simulation of the $e^+e^-\to K^+K^-$ process. 
The correction factor for the cut
$|\Delta\varphi| \leq \Delta\varphi_0$ was determined expanding the 
implied angular range by a factor of 2. 
 
\begin{figure}[htbp]
\includegraphics[width=0.6\textwidth]{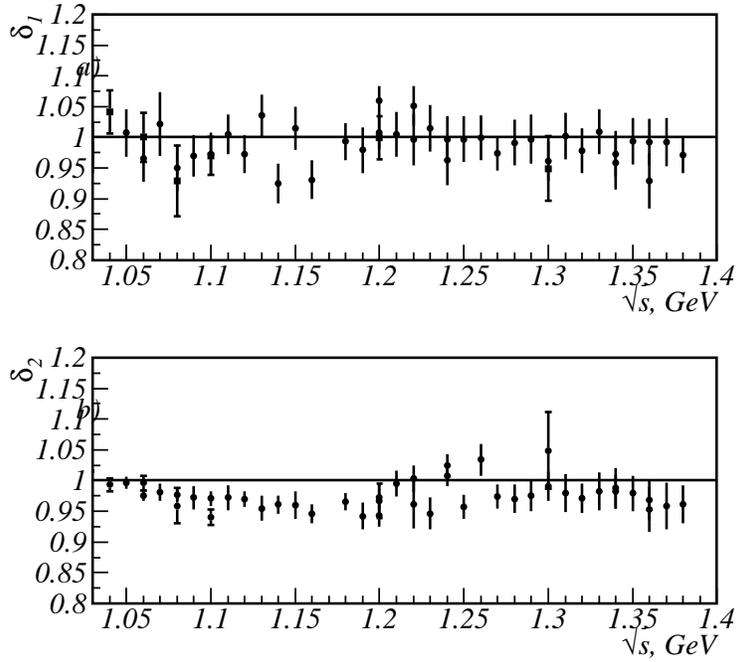}
\caption{ \label{deltatot}
The efficiency correction factors for all selection conditions except 
the cuts (\ref{kk}) (a), for the cuts (\ref{kk}) on the particles separation
parameters (b).
}
\end{figure}

The correction factor obtained for different energy points is shown in
Fig.\ref{deltatot}(a). The correction factor shown does not include the 
correction for the cuts (\ref{kk}), related to the particle 
separation parameters,
which is given separately in Fig.\ref{deltatot}(b). 
\noindent
The average accuracy of the correction factor determination 
is 4.4$\%$. The values of the corrected detection efficiencies
$\varepsilon_0(\sqrt{s})$  are listed in Table~\ref{tabcrs1}. 
The efficiency changes from 30\% to 50\% in the energy range
under study.
 
\section{ Cross section of the process $e^+e^- \to K^+K^-$}
The visible cross section $\sigma_{vis}$ for the process under study, 
directly observed in the experiment, is related to the Born cross section 
$\sigma_{K^+K^-}$ as follows
\begin{equation}
\label{si2}
 \sigma_{vis}(\sqrt{s}) =
    \int\limits_0^{1}
    \sigma_{K^+K^-}(\sqrt{s(1-z)})
    F(s,z)\varepsilon(\sqrt{s},z)dz,
\end{equation}
where $F(s,z)$ is a function describing a probability     
distribution of the energy fraction $z=2E_\gamma/\sqrt{s}$
taken away by the additional photon with energy   
$E_{\gamma}$ radiated from the initial           
state~\cite{fadin}. Formula  (\ref{si2}) can be rewritten in the 
traditional form
\begin{equation}
\label{si1}
\sigma_{vis}(\sqrt{s}) = \varepsilon_0(\sqrt{s})\sigma_{K^+K^-}(\sqrt{s})
(1+\delta(\sqrt{s})),
\end{equation}
where $\delta(\sqrt{s})$ is a radiative correction.
The following procedure was used to extract the experimental values of 
the Born cross section. The visible cross section 
for the $i$-th energy point is equal to 
\begin{equation}
\sigma_{vis,i}=\frac{N_{K^+K^-,i}}{IL_i},
\end{equation}
where $N_{K^+K^-,i}$ is the number of selected $e^+e^- \to K^+K^-$ events 
and $IL_i$ is an integrated luminosity. The measured energy
dependence of the visible cross section is approximated by 
a function calculated according to the formula (\ref{si2}) with the use
of several models for the Born cross section.
As a result of the approximation, the parameters of the model are 
determined and the following function is calculated
\begin{equation}
R(\sqrt{s})=\frac{\sigma_{vis}(\sqrt{s})}{\sigma_{K^+K^-}(\sqrt{s})}.
\end{equation}
Then the experimental Born cross section is determined through the formula 
\begin{equation}
\sigma_{K^+K^-,i}=\frac{\sigma_{vis,i}}{R(\sqrt{s_i})}.
\end{equation}
The model dependence of the result is estimated from its variation
under different models for the Born cross section. 

The Born cross section for the $e^+e^- \to K^+K^-$ was described
by the model of Vector Meson Dominance~\cite{achasov1,PDG}:
\begin{equation}
\label{sixth}
\begin{array}{l}
\sigma_{K^+K^-}(\sqrt{s}) =
\left|
\sum \limits_{V} A_V +
\sum \limits_{V^{\prime}} A_V^{\prime}
\right|^2,
\end{array}
\end{equation}
where
$V = \rho, \omega, \phi$, $V^{\prime} = \rho^{\prime}, \omega^{\prime}, 
\phi^{\prime} $.
The  $\rho$, $\omega$, $\phi$ mesons amplitudes were taken in the form:
\begin{equation}
\label{seventh}
\begin{array}{l}
A_V  = \sqrt{12\pi \displaystyle\frac
{ m_V^3 B_{V\to e^+e^-}\Gamma_V\Gamma_{V\to K^+K^-}(s)}
{s^{3/2}}} \cdot \displaystyle\frac{f_V}{D_V},\\
D_V(s) = m_V^2 - s - i \sqrt{s}\Gamma_V(s), \\
\Gamma_V(s) = \sum \limits_{f} \Gamma_{V\to f}(s),
~~\Gamma_{V\to f}(m_V^2) = \Gamma_V B_{V\to f},\\
\Gamma_{\rho\to K^+K^-}(s) = \Gamma_{\omega\to K^+K^-}(s) =
0.5\cdot\Gamma_{\phi\to K^+K^-}(s),
\end{array}
\end{equation}
where  $m_V$, $\Gamma_V$ are  mass and total width of the resonance,
$\Gamma_{V\to f}(s)$, $B_{V\to f}$ are  partial width and the branching 
ratio of the $V$ meson decay into the final state $f$, $f_V$ is the phase 
factor. Energy dependence of the resonance width was calculated 
taking into account the main decay modes. The parameters of the 
$\rho$, $\omega$ and $\phi$ mesons were taken from \cite{PDG}.

The amplitudes for the excited states $V^{\prime}$  were written as follows:
\begin{equation}
\label{eighth}
A_{V^{\prime}}  = 
\sqrt{\sigma_{V^{\prime}}\frac{W(s)}{W(m_{V^{\prime}}^2)}}
\frac{m_{V^{\prime}}\Gamma_{V^{\prime}}}{D_{V^{\prime}}}
f_{V^{\prime}},
\end{equation}
where $\sigma_{V^{\prime}}$ is the cross section of the process
$e^+e^- \to V^{\prime} \to K^+K^-$ at $s=m_{V^{\prime}}^2$.
\begin{equation}
\label{third}
\begin{array}{l}
W(s) =  q^3(s)/s^{5/2},\\
q(s) = \displaystyle\frac{\sqrt{s}}{2}(1-\frac{4m_K^2}{s})^{1/2}. 
\end{array}
\end{equation}
For masses and widths of $\omega^\prime, \phi^\prime$, 
the PDG values~\cite{PDG} were used.
The mass and width of the $\rho^\prime$, as well as
$\sigma_{V^{\prime}}$ 
and the relative phases $f_{V^{\prime}}$ were free parameters 
of the approximation.
The $\phi$ meson phase was 
varied in the limits $180^\circ \pm 30^\circ$ to estimate the model dependence
of the results. 
Besides, the approximation which takes into account the 
$\rho^{\prime\prime}$  and $\omega^{\prime\prime}$ meson contributions 
was performed. In this case the phase of the $\phi$ meson was fixed at 
$180^\circ$, while masses and widths of the excited states varied 
relative to the PDG data within their uncertainties. 
Depending on the model, $\chi^2/Nd$ varied in the limits of 
1.0-1.14.
\begin{figure}[htbp]
\includegraphics[width=0.68\textwidth]{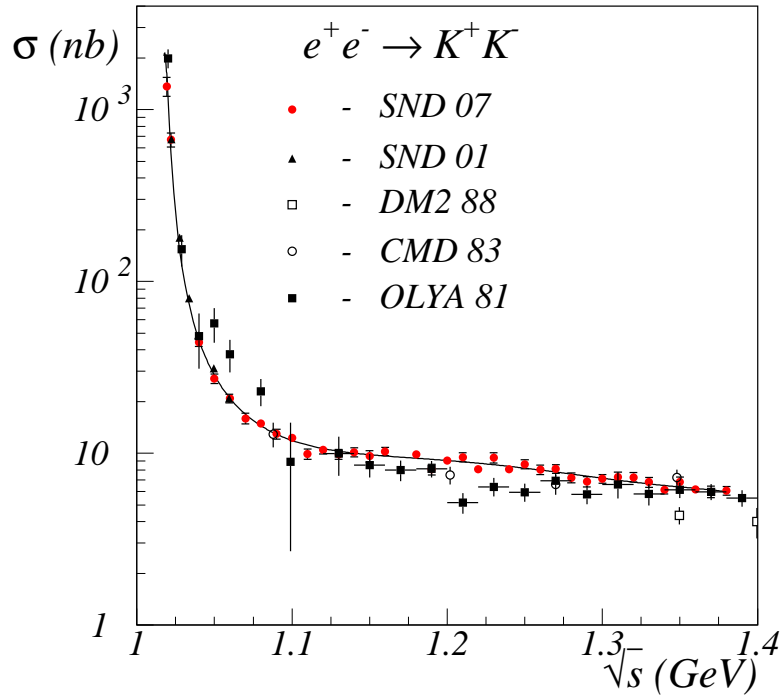}
\caption{ \label{crosskc}
The total cross section for the process $e^+e^- \to K^+K^-$ in the energy
range from 1.05 to 1.4~GeV obtained in different experiments.
SND~07 -- this work,
SND~01 -- \cite{ach1},
DM2~88 -- \cite{bisello88},
CMD~83 -- \cite{cmd1}.
OLYA~81 --\cite{olya}.
The solid line shows the result of the approximation.
}
\end{figure}
\begin{figure}[htbp]
\includegraphics[width=0.68\textwidth]{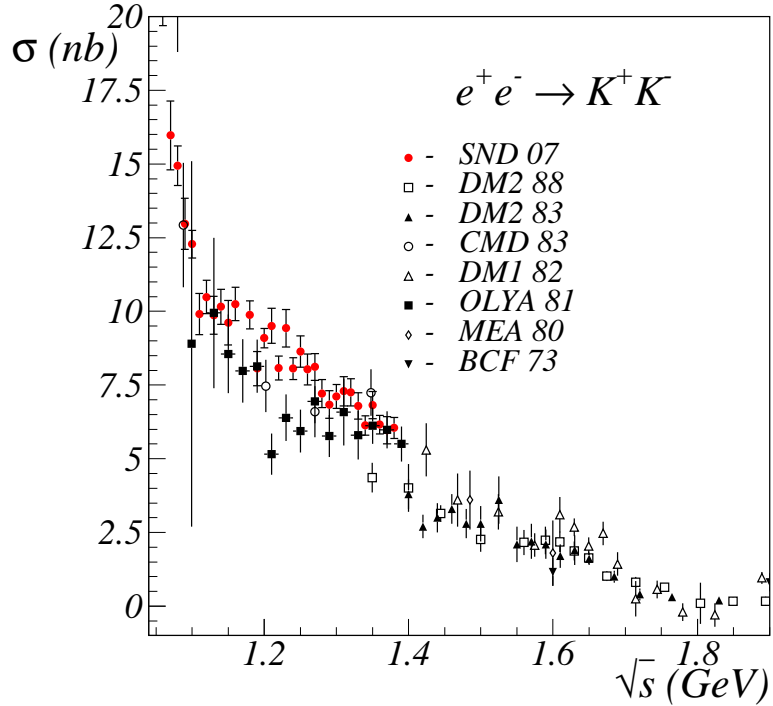}
\caption{ \label{crosskc1}
The $e^+e^- \to K^+K^-$ total cross section measured
in different experiments in the energy range from 1.05 to 2~GeV. 
SND~07  -- this work,
DM2~88 -- \cite{bisello88},
DM2~83 -- \cite{august},
CMD~83 -- \cite{cmd1},
DM1~81 -- \cite{dm1},
OLYA~81 -- \cite{olya},
MEA~80 -- \cite{mea},
BCF~73 -- \cite{bcf}.
}
\end{figure}

The obtained values of the Born cross section are shown in 
Figs.~\ref{crosskc},\ref{crosskc1} and listed in Table \ref{tabcrs1}. 
The table also gives the values of the charged kaon form factor which is 
related to the Born cross section as follows
\begin{equation}
\label{ninth}
\begin{array}{l}
\sigma_{K^+K^-}(\sqrt{s}) =\displaystyle \frac{\pi\alpha^2\beta^3}{3\cdot s}
|F_{K^+}(\sqrt{s})|^2, \\
\beta = \displaystyle \frac{2\cdot q(s)}{\sqrt{s}}.
\end{array}
\end{equation}
The radiative corrections calculated according to (\ref{si1}) are given as 
well. The average statistical error of the cross section measurement
is equal to 4.4~$\%$. An average systematic error is equal to 
5.2\% and includes inaccuracies in the determinations of the background 
(1.9\%), detection  efficiency (4.4\%), luminosity (2\%) and the model 
error (0.1\%).

\begin{table}[!ht]
\begin{center}
\caption{ \label{tabcrs1}
The $e^+e^- \to K^+K^-$ cross section ($\sigma_{K^+K^-}$) and 
charged kaon form factor ($|F_{K^+K^-}|$) measured in this experiment.
$E$ is the center-of-mass energy,
$IL$ is an integrated luminosity,
$\varepsilon_0$ is the detection efficiency,
$\delta$ is a radiative correction, 
$N(K^+K^-)$ is the number of selected $e^+e^- \to K^+K^-$ events.
}

{\scriptsize
\begin{tabular}{|c|c|c|c|c|c|c|}
\hline
     $E$, GeV & $IL$, nb$^{-1}$  & $\varepsilon_0$ & $(1+\delta)$
     & $N(K^+K^-)$ & $\sigma_{K^+K^-}$, nb
     & $|F_{K^+K^-}|^2$  \\
\hline
  1.04 &  69.75 & $0.357\pm0.013$&1.252 &$1347.4\pm36.7\pm12.3  $ & 
$44.30\pm1.66\pm1.88$&
$ 71.22 \pm 4.08 $\\[-1.25mm]                                 
  1.05 &  83.91 & $0.389\pm0.017$&1.164 &$1030.4\pm32.1\pm 9.0  $ & 
$27.12\pm1.01\pm1.34$&
$34.96 \pm  2.16 $\\[-1.25mm]
  1.06 & 279.8  & $0.446\pm0.014$&1.076 &$2832.3\pm53.2\pm 21.7 $ & 
$20.90\pm0.45\pm0.70$&
$22.46 \pm  0.91 $\\[-1.25mm]
  1.07 &  97.74 & $0.426\pm0.023$&1.026 &$682.0 \pm26.1\pm 8.2  $ & 
$15.96\pm0.64\pm0.96$&
$14.70 \pm  1.07 $\\[-1.25mm]
  1.08 & 550.47 & $0.427\pm0.016$&0.963 &$3392.0\pm58.2\pm 29.2 $ & 
$14.94\pm0.28\pm0.57$&
$12.06 \pm  0.53 $\\[-1.25mm]
  1.09 &  95.15 & $0.415\pm0.017$&0.901 &$465.5 \pm21.6\pm 6.5  $ & 
$12.97\pm0.61\pm0.60$&
$ 9.34 \pm  0.62 $\\[-1.25mm]
  1.10 & 388.1  & $0.474\pm0.012$&0.905 &$2042.0\pm45.2\pm 20.6 $ & 
$12.28\pm0.27\pm0.37$&
$ 7.99 \pm  0.30 $\\[-1.25mm]
  1.11 & 91.66  & $0.480\pm0.018$&0.889 &$387.0 \pm19.7\pm 2.5  $ & 
$ 9.91\pm0.51\pm0.47$&
$ 5.89 \pm  0.41 $\\[-1.25mm]
  1.12 & 231.26 & $0.511\pm0.017$&0.890 &$1101.6\pm33.2\pm 18.0 $ & 
$10.49\pm0.32\pm0.45$&
$ 5.76 \pm  0.30 $\\[-1.25mm]
  1.13 & 112.58 & $0.481\pm0.019$&0.884 &$473.0 \pm21.7\pm 2.9  $ & 
$ 9.87\pm0.46\pm0.45$&
$ 5.05 \pm  0.33 $\\[-1.25mm]
  1.14 & 188.44 & $0.473\pm0.018$&0.887 &$802.0 \pm28.3\pm 12.2 $ & 
$10.16\pm0.36\pm0.47$&
$ 4.87 \pm  0.28 $\\[-1.25mm]
  1.15 & 69.33  & $0.477\pm0.020$&0.888 &$281.7 \pm16.8\pm 4.6  $ & 
$ 9.62\pm0.57\pm0.48$&
$ 4.35 \pm  0.34 $\\[-1.25mm]
  1.16 & 211.48 & $0.468\pm0.017$&0.889 &$901.0 \pm30.0\pm 13.4 $ & 
$10.25\pm0.34\pm0.44$&
$ 4.39 \pm  0.24 $\\[-1.25mm]
  1.18 & 307.58 & $0.502\pm0.017$&0.891 &$1358.4\pm36.9\pm 14.8 $ & 
$ 9.88\pm0.27\pm0.38$&
$ 3.86 \pm  0.18 $\\[-1.25mm]
  1.19 & 172.29 & $0.444\pm0.020$&0.892 &$548.9 \pm23.4\pm 8.1  $ & 
$ 8.05\pm0.34\pm0.45$&
$ 3.02 \pm  0.21 $\\[-1.25mm]
  1.20 & 397.6  & $0.497\pm0.010$&0.894 &$1600.2\pm40.0\pm 15.0 $ & 
$ 9.09\pm0.23\pm0.22$&
$ 3.28 \pm  0.12 $\\[-1.25mm]
  1.21 & 151.45 & $0.481\pm0.020$&0.897 &$621.2 \pm24.9\pm 12.3 $ & 
$ 9.50\pm0.38\pm0.46$&
$ 3.32 \pm  0.21 $\\[-1.25mm]
  1.22 & 342.87 & $0.505\pm0.019$&0.898 &$1248.5\pm35.3\pm 21.8 $ & 
$ 8.07\pm0.23\pm0.32$&
$ 2.73 \pm  0.14 $\\[-1.25mm]
  1.23 & 140.75 & $0.444\pm0.021$&0.902 &$530.6 \pm23.0\pm 12.0 $ & 
$ 9.43\pm0.41\pm0.49$&
$ 3.10 \pm  0.21 $\\[-1.25mm]
  1.24 & 377.74 & $0.481\pm0.015$&0.903 &$1320.9\pm36.3\pm 22.5 $ & 
$ 8.06\pm0.22\pm0.30$&
$ 2.58 \pm  0.12 $\\[-1.25mm]
  1.25 & 209.00 & $0.428\pm0.018$&0.903 &$697.7 \pm26.4\pm 13.5 $ & 
$ 8.64\pm0.33\pm0.42$&
$2.69 \pm  0.17  $\\[-1.25mm] 
  1.26 & 162.90 & $0.496\pm0.022$&0.908 &$595.7 \pm24.4\pm 10.0 $ & 
$ 8.03\pm0.33\pm0.41$&
$ 2.45 \pm  0.16 $\\[-1.25mm]
  1.27 & 241.26 & $0.452\pm0.016$&0.907 &$802.0 \pm28.3\pm 16.0 $ & 
$ 8.14\pm0.29\pm0.36$&
$ 2.43 \pm  0.14 $\\[-1.25mm]
  1.28 & 228.98 & $0.472\pm0.021$&0.908 &$708.1 \pm26.6\pm 14.4 $ & 
$ 7.22\pm0.27\pm0.39$&
$ 2.11 \pm  0.14 $\\[-1.25mm]
  1.29 & 271.88 & $0.456\pm0.022$&0.909 &$770.8 \pm27.8\pm 15.4 $ & 
$ 6.85\pm0.25\pm0.40$&
$ 1.97 \pm  0.14 $\\[-1.25mm]
  1.30 & 271.14 & $0.453\pm0.017$&0.909 &$792.9 \pm28.2\pm 16.6 $ & 
$ 7.12\pm0.25\pm0.33$&
$ 2.01 \pm  0.12 $\\[-1.25mm]
  1.31 & 202.04 & $0.437\pm0.021$&0.910 &$585.6 \pm24.2\pm 11.4 $ & 
$ 7.30\pm0.30\pm0.40$&
$ 2.03 \pm  0.14 $\\[-1.25mm] 
  1.32 & 235.80 & $0.415\pm0.019$&0.910 &$645.8 \pm25.4\pm 12.7 $ & 
$ 7.26\pm0.29\pm0.37$&
$ 1.99 \pm  0.13 $\\[-1.25mm]
  1.33 & 292.78 & $0.420\pm0.020$&0.914 &$761.6 \pm27.6\pm 15.4 $ & 
$ 6.79\pm0.25\pm0.38$&
$ 1.84 \pm  0.12 $\\[-1.25mm] 
  1.34 & 438.67 & $0.456\pm0.017$&0.915 &$1121.8\pm33.5\pm 23.5 $ & 
$ 6.13\pm0.18\pm0.28$&
$ 1.64 \pm  0.09 $\\[-1.25mm]
  1.35 & 256.66 & $0.450\pm0.022$&0.912 &$718.7 \pm26.8\pm 21.5 $ & 
$ 6.82\pm0.26\pm0.39$&
$ 1.80 \pm  0.12 $\\[-1.25mm]
  1.36 & 625.16 & $0.432\pm0.018$&0.913 &$1524.2\pm39.0\pm 29.2 $ & 
$ 6.15\pm0.16\pm0.27$&
$ 1.61 \pm  0.08 $\\[-1.25mm]
  1.37 & 256.16 & $0.437\pm0.024$&0.914 &$610.8 \pm24.7\pm 17.2 $ & 
$ 5.97\pm0.24\pm0.39$&
$ 1.55 \pm  0.12 $\\[-1.25mm]
  1.38 & 479.75 & $0.424\pm0.019$&0.917	&$1127.3\pm33.6\pm 20.5 $ & 
$ 6.05\pm0.18\pm0.31$&
$ 1.56 \pm  0.09 $\\                       
\hline
\end{tabular}
}
\end{center}
\end{table}

\section{Discussion of the results}
Results of the measurement of the $e^+e^- \to K^+K^-$ cross section 
$\sigma_{K^+K^-}(\sqrt{s})$ in the energy range $\sqrt{s}$ = 1.04--1.38~GeV, 
presented in this work, are the most precise at the present time.
The obtained measurement errors, both statistical and systematic, are 2 
times less than achieved in the previous most precise experiment 
with the detector OLYA~\cite{olya}. Some deviations of the measured cross 
section from the results of \cite{olya} are observed.  In the energy range
1.05--1.08~GeV, the cross section obtained in \cite{olya} is 
approximately 60\% higher than our results (2 standard deviations), while 
in the energy range 1.15--1.25~GeV the results of \cite{olya} lay
systematically 25\% lower than those obtained in this work (3 standard 
deviations). On the other hand, our results agree well with the previous
SND measurements~\cite{ach1} in the energy range near the $\phi$ meson and
with results of ~\cite{cmd1} (Fig.\ref{crosskc}). 

\section{Conclusion}   
In this work the cross section of the process $e^+e^- \to K^+K^-$, and the 
charged kaon electromagnetic form factor, was measured in the energy range 
$\sqrt{s}$ = 1.04--1.38~GeV. The average statistical error of the 
measurement is 4.4\%, the systematic error is 5.2\%, which is approximately 
2 times better than that achieved in the previous most precise 
experiment~\cite{olya}. In general, the measured cross section
is consistent with 
the results of the previous experiments, but has better accuracy. 

We are grateful to S.Eidelman for his remarks and recommendations.

The work is supported in part by grants Sci.School-1335.2003.2,
RFBR 04-02-16184-a, RFBR 04-02-16181-a, RFBR 05-02-161250-a,
RFBR 06-02-16273-a,  RFBR 07-02-00104-a.

\end{document}